\documentstyle[aps,prb,multicol,graphicx]{revtex}

\begin{document}
\draft

\title{Effect of band filling in the Kondo lattice: A mean-field
approach}

\author{A. R. Ruppenthal, J. R. Iglesias, and M. A. Gusm\~ao}

\address{Instituto de F{\'\i}sica, Universidade Federal do Rio Grande
do Sul, Caixa Postal 15051, 91501-970 Porto Alegre, RS, Brazil}

\maketitle

\begin{abstract}
The usual Kondo-lattice, including an antiferromagnetic exchange
interaction between nearest-neighboring localized spins, is treated
here in a mean-field scheme that introduces two mean-field parameters:
one associated with the local Kondo effect, and the other related to
the magnetic correlations between localized spins.  Phases with
short-range magnetic correlations or coexistence between those and the
Kondo effect are obtained. By varying the number of electrons in the
conduction band, we notice that the Kondo effect tends to be
suppressed away from half filling, while magnetic correlations can
survive if the Heisenberg coupling is strong enough. An enhanced
linear coefficient of the specific heat is obtained at low
temperatures in the metallic state.
\end{abstract}


\begin{multicols}{2}

\section{Introduction} \label{sec:Intro}

The denomination {\it Kondo lattice\/}\cite{review} applies to systems
which are characterized by a lattice of localized magnetic moments
coexisting with a conduction band. Examples are found among Cerium
compounds, such as CeAl$_3$, CeCu$_6$, CeCu$_2$Si$_2$,
etc.,\cite{cerium} as well as some Uranium or other rare-earth
compounds.\cite{review} At high temperatures, the localized moments
behave essentially as independent impurities, similarly to what
happens in dilute alloys. At low temperatures, a coherent {\it
heavy-fermion\/} behavior is observed where the system resembles a
Fermi-liquid with enhanced values of parameters such as the
specific-heat constant $\gamma$ and the magnetic susceptibility
$\chi$. This is a scenario found in the Cerium compounds mentioned
above. However, deviations from this behavior are observed: some
systems can show a magnetically ordered ground state, as happens, for
example, in CeAl$_2$, CeB$_6$, and CePd$_2$Al$_3$, or become
superconducting at low temperatures, as is the case of
CeCu$_2$Si$_2$,\cite{stegl,aliev} CeCu$_2$Ge$_2$ at high
pressures,\cite{jaccard} and some Uranium compounds like
UPt$_3$,\cite{stewart} UBe$_{13}$,\cite{ott} and
URu$_2$Si$_2$,\cite{palstra} among others.  Besides that,
neutron-scattering experiments\cite{n1,n2,n3,n4} point to the presence
of strong short-range magnetic correlations in CeCu$_6$, CeInCu$_2$,
and CeRu$_2$Si$_2$, for which a Fermi-liquid ground state is stable,
but that are probably close to the conditions for a magnetic
instability.

This coexistence of magnetic correlations and Kondo effect was
recently discussed by Iglesias, Lacroix, and Coqblin\cite{turma}
within a mean-field approach that has points in common with a
slave-boson treatment by Coleman and Andrei.\cite{Coleman} Their
analysis was restricted to the case of a half-filled conduction band,
which is of relevance for Kondo insulators.\cite{insul} The role
played by the conduction-band filling is obviously very important for
metallic compounds that show heavy-fermion behavior, and to the
discussion of {\it underscreened\/} systems.\cite{screening}

In this paper we study the effect of band filling in the stability of
the Kondo state and short-range magnetic correlations in the Kondo
lattice using a mean-field approach closely related to the one
employed in Ref.~\onlinecite{turma}. Two mean-field parameters are
introduced, in connection with the local correlations generated by the
Kondo effect, and the non-local ones that indicate tendency towards
magnetic order. These quantities behave as order parameters of
thermodynamic phases where one of them or both are non-zero, while
both vanish in the high-temperature ``normal'' phase. Besides the
order parameters themselves, we calculate the free energy, the
specific heat, and the single-particle density of states. We focus on
the changes observed when one moves away from half-filling, as well as
the competition between correlations induced by the Kondo coupling and
those produced by a Heisenberg exchange term added to the original
Kondo-lattice model. When this interaction is strong, the system first
develops magnetic correlations, and only at a lower critical
temperature does the Kondo effect appear, depleting but not
suppressing the existing correlations. In contrast, when the
Heisenberg interaction is small compared to the local Kondo coupling,
the latter dominates, giving rise to a single phase where magnetic
correlations are mostly induced by the Kondo effect, and can even
change sign when the band filling is low enough. The specific heat
that we obtain in the metallic case shows an enhanced linear
coefficient, indicating that important qualitative aspects of the
physics of heavy-fermion systems are retained in the mean-field
approach.

In Sec.~\ref{sec:Model} we introduce the model Hamiltonian and the
fermionic representation of localized moments. In
Sec.~\ref{sec:Mean-field} we choose the relevant fields in terms of
which we rewrite the Hamiltonian, perform the mean-field decoupling,
find the energy eigenvalues, and introduce self-consistency
equations for the mean-field parameters, as well as relations to
obtain the physical quantities of interest. Our results are presented
in Sec.~\ref{sec:Results}, and a critical discussion of them appears
in Sec.~\ref{sec:Conclusions}.

\section{Model Hamiltonian} \label{sec:Model}

The usual Kondo-lattice Hamiltonian is
\begin{equation} \label{eq:kondo}
H = \sum_{{\bf k}\sigma} \varepsilon_{\bf k} n^c_{{\bf k}\sigma} -
J_{\text{K}} \sum_i {\bf s}^{}_i \cdot {\bf S}^{}_i \:,
\end{equation}
where ${\bf s}^{}_i$ stands for the total conduction-electron spin at
lattice site $i$, ${\bf S}^{}_i$ is the spin operator associated to
the localized moments, and the first term in the right-hand side
describes the conduction band in usual notation. We choose spin-1/2
localized moments, assigning them to ``$f$ electrons''. We also add a
Heisenberg-like interaction between nearest-neighbor localized
spins. The Kondo-lattice Hamiltonian then reads
\begin{eqnarray} \label{eq:model}
H &=& \sum_{{\bf k}\sigma} \varepsilon_{\bf k} n^c_{{\bf k}\sigma} + E_0
\sum_i n^f_{i\sigma} \nonumber \\ && \mbox{} 
- J_{\text{K}} \sum_i {\bf s}^{}_i \cdot {\bf
S}^{}_i - J_{\text{H}} \sum_{\langle ij \rangle} {\bf S}^{}_i \cdot {\bf
S}^{}_j \,,
\end{eqnarray}
The term with the factor $E_0$ is just a constant provided that
we remain in the subspace of unit occupation number for the $f$
electrons at every site, which is the constraint that ensures the
equivalence of Eqs.~(\ref{eq:kondo}) and~(\ref{eq:model}) if the
exchange interaction is also included in the original model.

The spin operators are rewritten in the usual fermionic representation
\begin{eqnarray} \label{eq:spins}
s_i^z = \frac{1}{2} \left( n^c_{i\uparrow} - n^c_{i\downarrow}
\right)\:, &&\qquad S_i^z = \frac{1}{2} \left( n^f_{i\uparrow} -
n^f_{i\downarrow} \right)\:, \nonumber \\ s_i^{+} =
c_{i\uparrow}^\dagger c_{i\downarrow} \:, &&\qquad S_i^{+} =
f_{i\uparrow}^\dagger f_{i\downarrow} \:, \\ s_i^{-} =
c_{i\downarrow}^\dagger c_{i\uparrow}  \:, &&\qquad
S_i^{-} = f_{i\downarrow}^\dagger f_{i\uparrow} \:. \nonumber
\end{eqnarray}
There is no unique way of rewriting the Hamiltonian in terms of fermion
operators. For instance, the terms involving the $z$ component of the
spins can be left as in Eq.~(\ref{eq:spins}), or the number operators
can be explicitly written as products of creation and annihilation
operators, whose ordering can then be altered using the fermionic
anticommutation relations. The final choice of form to write down
these terms will actually depend on the mean-field decoupling that
will be employed, as we will discuss in the next section.

\section{Mean-field scheme} \label{sec:Mean-field}

The first step in constructing the mean-field Hamiltonian is to choose
the relevant {\it fields\/}. The original Hamiltonian is naturally
written in terms of {\it spin fields\/}. However, if we just
introduced a mean-field decoupling of spin products we would end up
with a magnetic Hamiltonian. Since we intend to describe the Kondo
effect as well, we have to choose fields that couple $c$ and $f$
electrons at the same site, since these will tend to form local
singlets. We can also couple $f$ electrons from neighboring sites in
order to describe the short-range magnetic correlations that will
develop between the localized moments. A convenient way to do this is
to define the (Hermitian) fields
\begin{eqnarray} \label{eq:fields}
\hat\lambda^{}_{i\sigma} &\equiv & \frac{1}{2} \left(
c_{i\sigma}^\dagger f_{i\sigma}^{} + f_{i\sigma}^\dagger
c_{i\sigma}^{} \right) \:, \nonumber \\ && \\
\hat\Gamma^{}_{ij\sigma}
&\equiv & \frac{1}{2} \left( f_{i\sigma}^\dagger f_{j\sigma}^{} +
f_{j\sigma}^\dagger f_{i\sigma}^{} \right) \: , \nonumber
\end{eqnarray}
where $\sigma$ labels the spin orientation (up or down) with respect
to the $z$ axis, and $i$ and $j$ are nearest-neighbor sites in the
definition of $\hat\Gamma^{}_{ij\sigma}$.  It is straightforward to
write down the contributions of transversal spin components in terms
of these fields. We have
\begin{eqnarray} \label{eq:sxy}
s_i^x S_i^x + s_i^y S_i^y &=& - \sum_\sigma
\hat\lambda^{}_{i\sigma} \hat\lambda^{}_{i,-\sigma} \:, \nonumber \\
&&\\
S_i^x S_j^x + S_i^y S_j^y &=& - \sum_\sigma
\hat\Gamma^{}_{ij\sigma} \hat\Gamma^{}_{ij,-\sigma} \:,\nonumber
\end{eqnarray}
where we have made explicit usage of the constraint $n_i^f=1$. 
Products involving the $z$ component of the spins are slightly more
complicated to deal with, and sometimes these are decoupled as spin
fields.\cite{turma} However, this breaks the spin symmetry, and
results in the stability of a magnetic state, with total suppression
of the Kondo effect, unless the local average magnetic moment is
forced to be zero. Hence, we will avoid this procedure, using the same
kind of representation for all spin components. After some algebraic
manipulations, we obtain
\begin{eqnarray} \label{eq:szsz}
s_i^z S_i^z &=& \frac{1}{4} \left( n_i^c + n_i^f \right) -
\frac{1}{4} n_i^c n_i^f - \sum_\sigma
\hat\lambda^{2}_{i\sigma}  \:, \nonumber \\ && \\
S_i^z S_j^z &=& \frac{1}{4} \left( n_i^f + n_j^f \right) - 
 \frac{1}{4} n_i^f n_j^f - \sum_\sigma
\hat\Gamma^{2}_{ij\sigma} \:. \nonumber
\end{eqnarray}
Due to the constraint $n_i^f=1$ at all sites, the terms containing
number operators in Eqs.~(\ref{eq:szsz}) will yield a constant shift
of the energies (after summation over the lattice sites). This terms
can then be dropped, leaving the Kondo and Heisenberg parts of the
Hamiltonian respectively written as
\begin{equation} \label{eq:HK}
H_{\text{K}} = J_{\text{K}} \sum_{i\sigma} \left(
\hat\lambda_{i\sigma} + \hat\lambda_{i,-\sigma} \right)
\hat\lambda_{i\sigma} \:,
\end{equation}
and
\begin{equation} \label{eq:HH}
H_{\text{H}} = J_{\text{H}} \sum_{\langle ij\rangle \sigma} \left(
\hat\Gamma_{ij\sigma} + \hat\Gamma_{ij,-\sigma} \right)
\hat\Gamma_{ij\sigma} \:.
\end{equation}

We now proceed with a standard mean-field decoupling of the operator
products in these Hamiltonians, introducing the mean-fields
$\lambda_{i\sigma} \equiv \langle \hat\lambda_{i\sigma} \rangle$ and
$\Gamma_{ij\sigma} \equiv \langle \hat\Gamma_{ij\sigma} \rangle$. 
From Eqs.~(\ref{eq:HK}) and~(\ref{eq:HH}) ones sees that each field
couples with a spin-independent quantity. Thus, there will be no
breakdown of spin symmetry, i.\ e., no magnetic states. Taking this
into account, together with translational invariance, we can simplify
the notation, using  $\lambda_{i\sigma} \equiv \lambda$ and
$\Gamma_{ij\sigma} \equiv \Gamma$. 

We are now in a position to put all contributions together: the
noninteracting terms of Eq.~(\ref{eq:model}) and the mean-field
versions of the interaction Hamiltonians (\ref{eq:HK})
and~(\ref{eq:HH}). Reverting to the fermionic representation, and
adding a chemical potential $\mu$ for the conduction electrons, we can
write the complete mean-field Hamiltonian
\begin{eqnarray} \label{eq:HMF}
H_{}^{\text{MF}} &=& \sum_{{\bf k}\sigma} (\varepsilon_{\bf k}-\mu)
n^c_{{\bf k}\sigma} + E_0 \sum_i n^f_{i\sigma} \nonumber \\ && \mbox{}
+ 2 J_{\text{K}} \lambda \sum_{i\sigma} \left( c_{i\sigma}^\dagger
f_{i\sigma}^{} + f_{i\sigma}^\dagger c_{i\sigma}^{} \right) +
\bar{E}_{\text{K}} \nonumber \\ && \mbox{} 
+ 2 J_{\text{H}} \Gamma \sum_{\langle ij\rangle \sigma} \left(
f_{i\sigma}^\dagger f_{j\sigma}^{} + f_{j\sigma}^\dagger f_{i\sigma}^{}
\right) + \bar{E}_{\text{H}} \:,
\end{eqnarray}
with 
\begin{equation} \label{eq:Ebarsfinal}
\bar{E}_{\text{K}} = - 4 N J_{\text{K}} \lambda^2 \:, \quad
\bar{E}_{\text{H}} = - 2 z N J_{\text{H}} \Gamma^2 \:,
\end{equation}
where $z$ represents the coordination number of a site in the lattice,
and $N$ is the total number of sites.

We will choose the conduction band to be a tight-binding band of width
$2W$ (with nearest-neighbor hopping only) in a (hyper)cubic lattice in
$d$ dimensions, so that we can write
\begin{equation} \label{eq:band}
\varepsilon_{\bf k} = - \frac{W}{d} \sum_\mu \cos(k_\mu a) \:,
\end{equation}
where $\mu$ labels the wavevector components, and $a$ is the lattice
parameter. With this choice, the term with $J_{\text{H}}$ in
Eq.~(\ref{eq:HMF}) will have the same ${\bf k}$-dependence as the
conduction band, except for the band width. We see, thus, that the
mean-field decoupling introduces a non-zero band width to the $f$
electrons. At this point, we have to replace the constraint $n_i^f=1$
with the much weaker constraint $\langle n_i^f \rangle = 1$. Hence,
the virtual charge fluctuations introduced by the fermion
representation of the localized spins have been ``put in the mass
shell'' by the mean-field decoupling. Of course, this real charge
fluctuations are spurious, but the method relies on the expectation
that the physics of the system will be reasonably preserved on the
phase-space surface where the constraint $\langle n_i^f \rangle = 1$
holds. In order to improve on this point one would have
to envisage better ways to take into account the constraint of single
occupancy on the $f$ levels at each site.

Leaving aside, for the moment, the constant terms ${E}_{\text{K}}$ and
${E}_{\text{H}}$, the Hamiltonian~(\ref{eq:HMF}) is easily
diagonalized, yielding the energy eigenvalues\cite{turma}
\begin{eqnarray} \label{eq:eigen}
E_{\bf k}^{\pm} &=& \frac{1}{2} \bigg[ \,\varepsilon_{\bf k} (1+B) +
E_0 -\mu \nonumber \\ && \mbox{~} 
\pm \sqrt{\left[ \varepsilon_{\bf k}
(1-B) - E_0 -\mu \right]^2 + 16J_{\text{K}}^2 \lambda^2} \; \bigg] \:,
\end{eqnarray}
where $B \equiv -4dJ_{\text{H}}\Gamma/W$. These energies are
spin-degenerate, as we remarked before. With them and the
corresponding eigenvectors we can compute any average of relevant
physical quantities in the system. The chemical potential $\mu$ is
determined by the equality $\langle n_{\bf k}^c \rangle = n$, where
$n$ is the (chosen) density of conduction electrons. $E_0$ plays the
role of a chemical potential for the $f$ electrons, being fixed by the
condition $\langle n_{\bf k}^f \rangle = 1$. The mean-field parameters
$\lambda$ and $\Gamma$ are determined through the selfconsistency
equations
\begin{eqnarray} \label{eq:selfcons}
\lambda &=& \frac{1}{2N}\sum_{\bf k} \left\langle c_{{\bf
k}\sigma}^\dagger f_{{\bf k}\sigma}^{} + f_{{\bf k}\sigma}^\dagger
c_{{\bf k}\sigma}^{} \right\rangle \:, \nonumber \\ && \\ 
\Gamma &=& - \frac{1}{WN} \sum_{\bf k} \varepsilon_{\bf k} \langle
n^f_{{\bf k} \sigma} \rangle \:. \nonumber 
\end{eqnarray}

We also check the self-consistency process by evaluating the
free-energy, which can be written as
\begin{eqnarray} \label{eq:freen}
F &=&  - 2T\sum_{{\bf k}, \alpha=\pm}
\log\left[ 1 + e^{-E_{\bf k}^\alpha/T}\right] 
\nonumber \\ && \mbox{}
+ \bar E_{\text{K}} + \bar E_{\text{H}} - (E_0 - \mu n)N \:.
\end{eqnarray}
where $T$ is the temperature (in energy units). Finally, we can
calculate the average internal energy
\begin{eqnarray} \label{eq:energ}
E = 2\sum_{{\bf k}, \alpha=\pm} E_{\bf k}^\alpha f(E_{\bf k}^\alpha) +
\bar E_{\text{K}} + \bar E_{\text{H}} - (E_0 - \mu n)N \:,
\end{eqnarray}
where $f(\varepsilon)$ stands for the Fermi function. The last term in
Eqs.~(\ref{eq:freen}) and~(\ref{eq:energ}) is needed to compensate for
the fact that $\mu$ and $E_0$ are already included in the
Hamiltonian. The specific heat can now be obtained as $c^{}_V=\partial
E/\partial T$, and we can check for an enhanced linear coefficient
$\gamma$, which is a signature of heavy-fermion behavior.

\section{Results} \label{sec:Results}

In order to be close to real systems, we considered a
three-dimensional simple-cubic lattice ($z=6$). However, given that
all wave-vector dependence occurs through $\varepsilon_{\bf k}$, we
turn all ${\bf k}$ sums into integrals over the conduction-band
energies, for which we use a semi-elliptical density of states of the
form 
\begin{equation} \label{eq:dens}
{\cal D}(\varepsilon) = \left\{ \begin{array}{lc} \frac{3}{4} (1-
\varepsilon^2)\;, \quad & -1 \le \varepsilon \le 1 \:; \\ 0 \:, \quad
& \mbox{otherwise} \:. \end{array} \right.
\end{equation}
Here we have set $W=1$, so that all energies are measured in units of
the half band width. We choose both $J_{\text{K}}$ and $J_{\text{H}}$ to be
negative, which corresponds to the usual Kondo coupling, and
antiferromagnetic Heisenberg interactions between nearest-neighboring
localized spins. The latter would be naturally interpreted as
generated by a superexchange mechanism, and could also mimic the RKKY
interaction in the mean-field Hamiltonian, although this is a
debatable issue: we will see below that there exist {\it induced\/}
spin correlations due to the Kondo coupling alone, and these are
more closely related to the RKKY mechanism.

\narrowtext
\begin{figure}
\begin{center}
\includegraphics[width=8.0cm,angle=0]{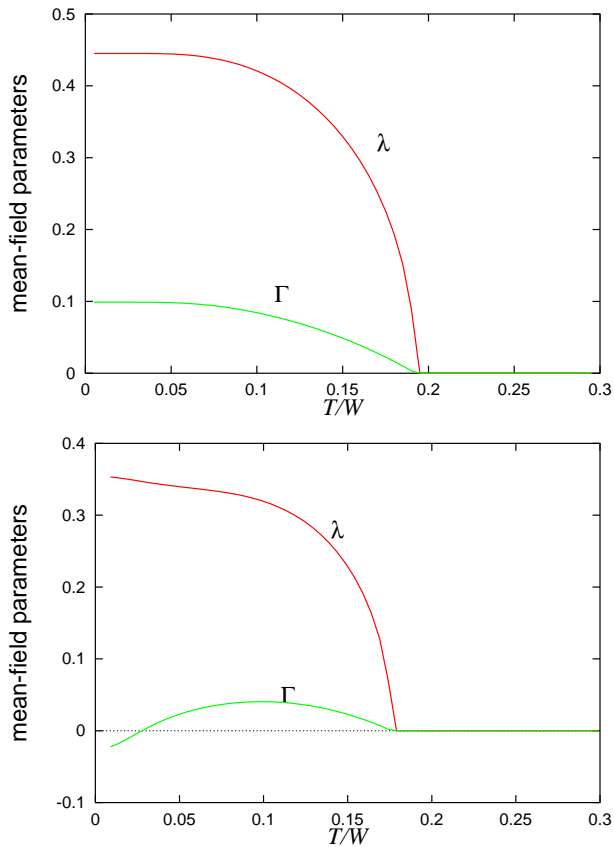}
\end{center}
\caption{Mean-field parameters $\lambda$ and $\Gamma$ as functions of
temperature for $n=1$ (top) and $n=0.6$ (bottom), in the case where
the Heisenberg interaction is small compared to the Kondo coupling:
$J_{\text{K}}/W=-0.5$, $J_{\text{H}}/W=-0.1$.}
\label{fig:lGxTJHsm}
\end{figure}

\begin{figure}
\begin{center}
\includegraphics[width=8.5cm,angle=0]{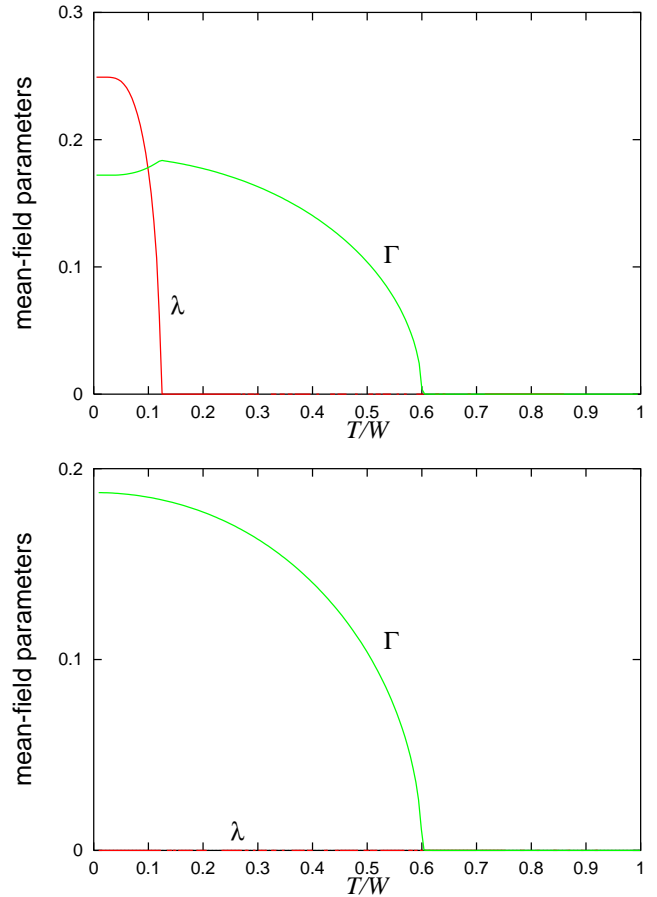}
\end{center}
\caption{Same as in Fig.\ \protect\ref{fig:lGxTJHsm}, but with a large
Heisenberg interaction: $J_{\text{K}}/W=-0.5$, $J_{\text{H}}/W=-1.0$.}
\label{fig:lGxTJHla}
\end{figure}

After numerically solving the set of self-consistency equations, we
find that both $\lambda$ and $\Gamma$ behave as typical mean-field
order parameters, which vanish at high temperatures, and become
non-zero through second-order phase transitions at well defined
critical temperatures. We identify the region where $\lambda \ne 0$
with the Kondo regime, and the corresponding critical temperature is
interpreted as the Kondo temperature ($T_{\text{K}}$) for the
lattice. We call $T_{\text{Corr}}$ the temperature below which
short-range magnetic correlations appear ($\Gamma \ne 0$).  When
$|J_{\text{H}}| << |J_{\text{K}}|$, both transitions occur at the same
temperature, while $|J_{\text{H}}|$ comparable to or larger than
$|J_{\rm K}|$ gives $T_{\text{Corr}} > T_{\text{K}}$, as obtained
earlier for the half-filled case.\cite{turma} We show a typical
example of this behavior in Figs.~\ref{fig:lGxTJHsm} and
\ref{fig:lGxTJHla}, where we compare the results for a half-filled and
a less-than-half-filled conduction band.  (Due to particle-hole
symmetry, the results are actually symmetric around $n=1$.) 

Departure from the half-filled situation causes a reduction of the
critical temperatures, except for $T_{\text{Corr}}$ in the
large-$|J_{\text{H}}|$ regime. Although the sign of $\Gamma$ does not
have a direct physical meaning, since spin correlations are related to
the square of the operator associated to $\Gamma$ according to
Eq.~(\ref{eq:szsz}), we believe that the change of sign observed in
Fig.~\ref{fig:lGxTJHsm} for the case of low band filling indicates
that the nearest-neighboring spin correlations have become
ferromagnetic, as expected in this limit. This is only observed at low
temperatures, where the Kondo effect is large, and the induced part of
the correlations dominate. In contrast, close to $T_{\text{K}}$,
$\lambda$ is small, and the antiferromagnetic correlations due to the
Heisenberg interaction dominate. In order to check this, we would have
to calculate the average $\langle {\hat\Gamma}_{ij\sigma}^2 \rangle$
that appears in the spin-spin correlations,
Eq.~(\ref{eq:szsz}). Unfortunately, this is related to averages of
products of number operators in each site, which are non-trivial in
spite of the non-interacting nature of the effective Hamiltonian. More
specifically, these products involve number operators at displaced
wavevectors, which prevents the transformation of momentum summations
into energy integrals. Notice, however, that when $\Gamma$ goes to
zero, spins in different sites become decoupled in the mean-field
Hamiltonian. Then, their correlations go to zero, which should happen
when they change from antiferro- to ferromagnetic.

\begin{figure}
\begin{center}
\includegraphics[totalheight=7.5cm,angle=-90]{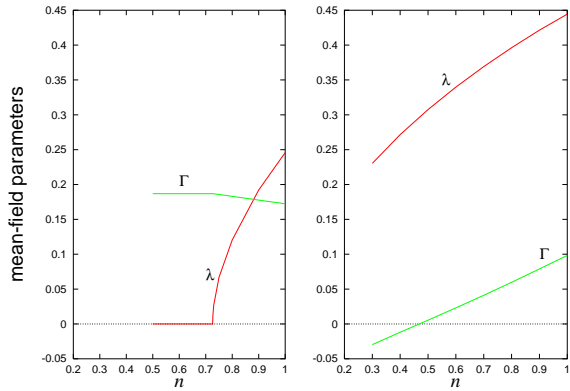}
\end{center}
\caption{Variation of the mean-field parameters with band filling at
fixed temperature ($T/W=0.05$). In the left panel we have dominant
magnetic correlations ($J_{\text{K}}/W=-0.5$,
$J_{\text{H}}/W=-1.0$), while in the right panel the Kondo effect
dominates ($J_{\text{K}}/W=-0.5$, $J_{\text{H}}/W=-0.1$).}
\label{fig:lGxn}
\end{figure}
\begin{figure}
\begin{flushleft}
\includegraphics[totalheight=10.2cm,angle=0]{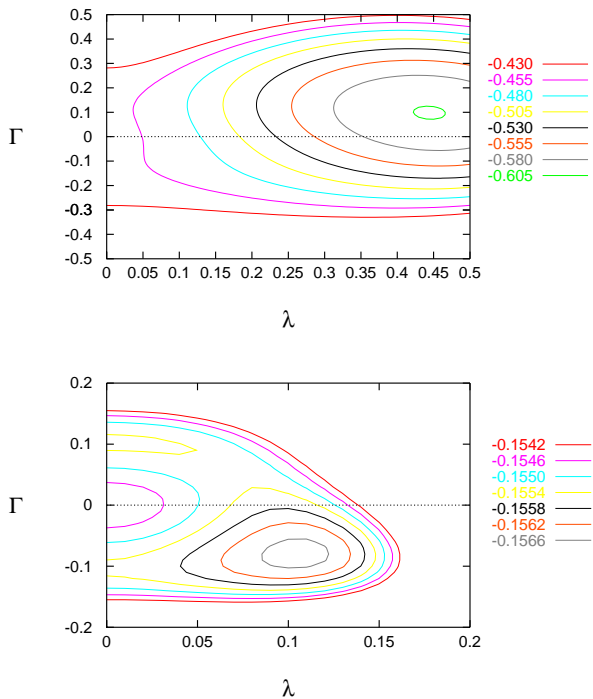}
\end{flushleft}
\caption{Equal free-energy contours at $n=1$ (top) and $n=0.1$
(bottom) for $|J_{\text{H}}|$ small, showing the complex topolgy that
arises at low filling. The values of the free energy for each
contour, quoted on the right, are in units of the half band width.}
\label{fig:freen}
\end{figure}

 In Fig.~\ref{fig:lGxn} we plot the mean-field parameters as functions
of the band filling for a fixed (low) temperature (T/W=0.05). For
large $|J_{\text{H}}|$ (left panel), the Heisenberg coupling
dominates, keeping $\Gamma$ almost unchanged, while $\lambda$ is
strongly reduced as we move away from $n=1$, since we have an
underscreening situation. This picture is completely changed in the
small-$|J_{\text{H}}|$ regime (right panel).  Now both parameters are
reduced as one moves away from half-filling.  One sees that $\Gamma$
is now equally sensitive to $n$, which suggests that in this regime
the part of the inter-site correlations {\it induced\/} by the
conduction electrons is dominant. It eventually yields ferromagnetic
spin correlations, as discussed above. We checked that a similar
behavior is observed if we fix $J_{\text{H}}=0$, when all spin
correlations are induced by the conduction electrons.

We have cut the curves in the right panel of Fig.~\ref{fig:lGxn} for
small $n$ because the convergence of our numerical calculations, which
involve iterating the set of self-consistency equations for the
mean-field parameters and occupation numbers, becomes very delicate in
that region.  In order to see what is happening, we evaluate the free
energy $F$ [Eq.~(\ref{eq:freen})] for fixed $\lambda$, $\Gamma$, $n$,
and $T$, adjusting only $\mu$ and $E_0$. Typical contour plots of $F$
at low temperature as a function of $\lambda$ and $\Gamma$, for
half-filling and for $n$ very small, in the small-$|J_{\text{H}}|$
regime, are shown in Fig.~\ref{fig:freen}. We can see that the minimum
at finite $\lambda$ and $\Gamma$ that exists at half-filling is
displaced towards small values of $\lambda$ and negative values of
$\Gamma$ as $n$ is reduced. Eventually, a new minimum appears for
$\Gamma > 0$, since the energy eigenvalues, Eq.~(\ref{eq:eigen}), are
symmetric with respect to the sign of $\Gamma$ when $\lambda=0$. Thus,
when $\lambda$ approaches zero (as $n$ is reduced) we have two shallow
minima separated by a very broad and low maximum at
$\lambda=\Gamma=0$, which causes problems to the numerical iteration
of the self-consistency equations since distant points in the
$\lambda$-$\Gamma$ space are very close in free-energy.

\begin{figure}
\begin{center}
\includegraphics[totalheight=7.0cm,angle=-90]{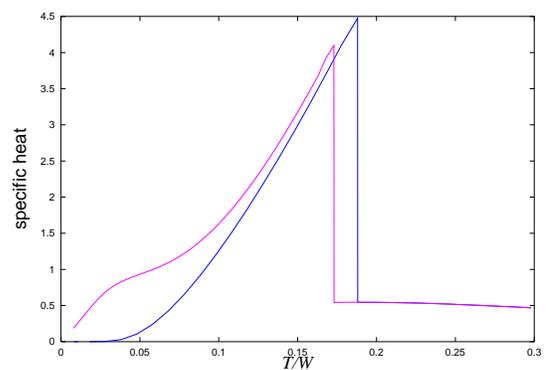}
\end{center}
\caption{Variation of the specific heat (per unit cell) with
temperature for $n=1$ (continuous) and $n=0.6$ (dashed), when the
Heisenberg interaction is weak ($J_{\text{K}}/W=-0.5$,
$J_{\text{H}}/W=-0.1$). Notice the insulating and metallic behavior at
low temperatures in each case, and the lambda-shaped transition at the
point where both the Kondo effect and magnetic correlations
disappear.}
\label{fig:spheatJHsm}
\end{figure}

\begin{figure}
\begin{center}
\includegraphics[totalheight=7.5cm,angle=-90]{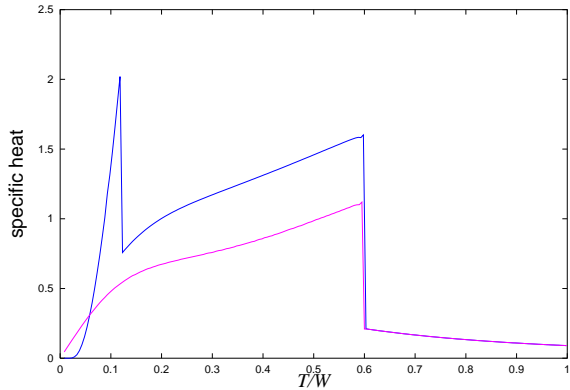}
\end{center}
\caption{Same as in Fig.\ \protect\ref{fig:spheatJHsm}, but for strong
Heisenberg coupling ($J_{\text{K}}/W=-0.5$, $J_{\text{H}}/W=-1.0$),
showing the transitions at different temperatures.}
\label{fig:spheatJHla}
\end{figure}

\begin{figure}
\begin{center}
\includegraphics[width=8.5cm,angle=0]{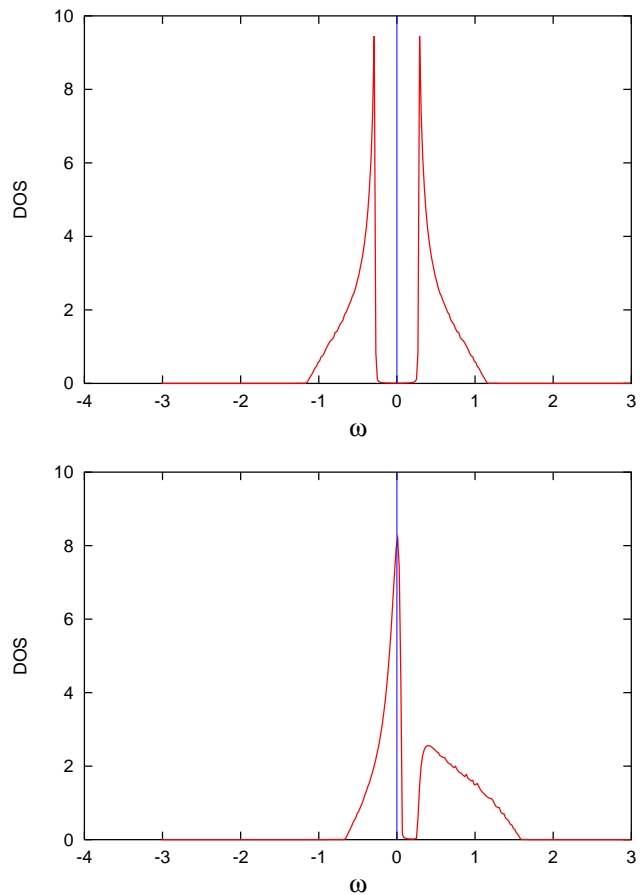}
\end{center}
\caption{Total single-particle density of states at $T/W=0.01$, for
$n=1$ (top) and $n=0.6$ (bottom), with $J_{\text{K}}/W=-0.5$,
$J_{\text{H}}/W=-0.1$. The frequencies are measured in
units of $W$. Notice that the Fermi level (vertical dashed line) lies in
the gap at half-filling, and inside a high peak for $n<1$.}
\label{fig:DOS}
\end{figure}

Our results for the specific heat as a function of temperature are
presented in Figs.~\ref{fig:spheatJHsm} and \ref{fig:spheatJHla}. The
phase transitions appear in the usual ``lambda shape'' characteristic
of mean-field approximations. There are two jumps in the case of large
$|J_{\text{H}}|$, at the temperatures where the Kondo effect and the
magnetic correlations disappear. These points coincide when
$|J_{\text{H}}|$ is small. We can also see the existence of a gap at
the Fermi level for the half-filled band, yielding an exponentially
vanishing specific heat for $T \to 0$. In contrast, away from
half-filling one sees a linear regime at low temperatures. The linear
coefficient obtained for $n=0.6$, $|J_{\text{K}}|=0.5$, and
$|J_{\text{H}}| = 0.01$ is $\gamma \sim 10^{} \gamma_0$, where
$\gamma_0$ is the corresponding value in the absence of
interactions. We have, thus, some enhancement of $\gamma$, although
the true heavy-fermion limit ($\gamma \sim 10^2 \gamma_0$) is not
reached. Such an enhancement should correspond to a high density of
states (DOS) at the Fermi level. We calculated the DOS as
\begin{eqnarray} \label{eq:DOS}
\rho(\omega) = -\frac{1}{\pi} \sum_{\alpha=\pm}  \sum_{{\bf k}\sigma}
{\rm Im} \,G_{{\bf k}\sigma}^\alpha (\omega + i0^+) \:,
\end{eqnarray}
in terms of the one-particle Green's functions
\begin{eqnarray} \label{eq:GFs}
G_{{\bf k}\sigma}^\alpha (z) = \frac{1}{z - E_{\bf k}^\alpha} \:.
\end{eqnarray}
The DOS depends on temperature through $\lambda$ and $\Gamma$, which
appear in the energies $E_{\bf k}^\alpha$. We plot the
low-temperature DOS in Fig.~\ref{fig:DOS}, for both $n=1$ and $n=0.6$,
in the small-$|J_{\text{H}}|$ regime. We can see that the Fermi level
($\omega=0)$ falls in the gap for $n=1$, while it is located at a
point corresponding to a very high density of states away from half
filling.  The gap that is observed immediately above the Fermi level
in the latter case explains the reduction (or at least levelling off)
of the specific heat just above the linear region in
Fig.~\ref{fig:spheatJHsm}. As the temperature is increased, the gap in
the DOS as well as the peaks around it are narrowed until a
superposition of a free conduction band and a localized $f$-level is
recovered above $T_{\text{K}}$.

\section{Conclusions} \label{sec:Conclusions}

We have presented here a detailed mean-field analysis of the Kondo
lattice, emphasizing the effects of conduction-band filling in the
properties of the system. The regime where the Kondo effect is
observed, and the one in which short-range magnetic correlations are
present appear as thermodynamic phases with well defined critical
temperatures. These phases are characterized by non-zero values of the
mean-field parameters associated to local correlations between
localized and conduction electrons ($\lambda$), and between localized
electrons on nearest-neighboring sites ($\Gamma$). We find a regime in
which the Kondo effect and short-range magnetic correlations coexist,
in agrrement with experimental observations in some Ce
compounds.\cite{n1,n2,n3,n4} The main consequence of changing the
electronic density in the conduction band with respect to half-filling
is a tendency to suppress the Kondo effect. Magnetic correlations are
equally suppressed if a direct Heisenberg interaction between
nearest-neighboring localized spins is weak, while they are almost
insensitive to the band filling for strong Heisenberg interactions. In
the latter case the critical temperatures for the two order parameters
are different, and one notices a partial suppression of magnetic
correlations in the region where the Kondo effect is observed. The
phase transitions appear as lambda-shaped peaks in the specific heat,
which also shows an enhanced linear coefficient in the metallic
situation ($n \ne 1$), corresponding to an enhanced density of states
at the Fermi level.

To a good extent, the physics of Kondo systems is recovered here, at
least qualitatively, both for insulators ($n=1$) and conductors ($n
\ne 1$).   A weak point of this treatment is the
impossibility to describe magnetically ordered states. Here one would
have to go back to the choice of relevant fields, trying to write down
a Hamiltonian in which both Kondo and spin fields were present before
the mean-field decoupling. It would be interesting to improve the
method along these lines without loosing its formal simplicity. This
simplicity constitutes the main quality of this approach, allowing 
evaluation of many relevant physical quantities in an unrestricted
range of parameters and conditions, and yielding qualitative
understanding of the behavior of the system as a whole.

\acknowledgments
This work was supported by CNPq (Brazil) and Brazil-France
agreement CAPES-COFECUB 196/96. M. A. G. benefited from the grant
FINEP-PRONEX no.\ 41.96.0907.00 (Brazil). We are indebted to
B. Coqblin and C. Lacroix for enlightening discussions.

\end{multicols}

\end{document}